\documentclass[conference]{IEEEtran}
		
\newcommand{\E}{\mathrm{E}}

\usepackage{epsf,psfrag,amssymb,amsfonts,color,cite,fancybox}
\usepackage[mathscr]{eucal}
\usepackage[dvips]{graphicx}
\usepackage{caption}
\usepackage{subcaption}
\usepackage{ifpdf}
\usepackage{longtable}	
\usepackage{multicol}
\usepackage{float}
\usepackage{epstopdf}
\usepackage{bm}
\usepackage{multirow}
\usepackage[scientific-notation=true]{siunitx}

\ifCLASSINFOpdf

\else

\fi

\usepackage[cmex10]{amsmath}
\begin{document}


\title{PMU-Based Estimation of Dynamic State Jacobian Matrix} 

\author{\IEEEauthorblockN{Xiaozhe Wang}
\IEEEauthorblockA{Department of Electrical and  Computer Engineering\\
McGill University\\
Montreal, Quebec H3A 0G4, Canada\\
Email: xiaozhe.wang2@mcgill.ca}
\and
\IEEEauthorblockN{Konstantin Turitsyn}
\IEEEauthorblockA{Department of Mechanical Engineering\\
Massachusetts Institute of Technology\\
Cambridge, MA 02139, USA\\
Email: turitsyn@mit.edu}
}

\maketitle

\begin{abstract}
In this paper, a hybrid measurement and model-based method is proposed which can estimate the dynamic state Jacobian matrix in near real-time. The proposed method is computationally efficient and robust to the variation of network topology. A numerical example is given to show that the proposed method is able to provide good estimation for the dynamic state Jacobian matrix and is superior to the model-based method under undetectable network topology change. The proposed method may also help identify big discrepancy in the assumed network model.
\end{abstract}


\IEEEpeerreviewmaketitle

\section{Introduction}
Security assessment such as stability analysis generally requires repeated computation based on the full nonlinear power system model, leading to huge computational efforts \cite{Chiang:book}. In addition, the security analysis strongly depends on an accurate network model, which may not be available due to communication failures \cite{Chen:2014}. These factors pose great challenges to online security analysis. Until recently, the deployment of phasor measurement units (PMUs) provides a great opportunity for the development of measurement-based security analysis methods in power systems \cite{Wangxz:2015}-\cite{Zhou:2009PES}.

In this paper, we propose a hybrid measurement and model-based method to estimate the dynamic state Jacobian matrix in near real-time, which provides invaluable information for various security analysis. Conventionally, the Jacobian matrix can be constructed based on state estimation results provided that an accurate dynamic model and network parameter values are available \cite{Abur:book}, which unfortunately is not the case in practice. As a result, the conventional method may give rise to imprecise estimations. In contrast, the proposed hybrid method does not depend on network model, and thus can work as a robust alternative to traditional state estimation-based approaches when uncertainty in network topology is an issue. In addition, the proposed method may also help system operators identify discrepancies in the assumed network model.


The estimated dynamic state Jacobian matrix can be utilized in various applications such as online oscillation analysis, stability monitoring and generation re-dispatch. Due to page limit, we have to present the detailed applications of the estimated matrix in a separate paper.

\section{estimating dynamic state jacobian matrix}\label{sectionmodel}
We consider the general power system dynamic model: 
\begin{eqnarray}
\dot{\bm{x}}&=&\bm{f}({\bm{x},\bm{y}})\label{fast ode}\\
\bm{0}&=&\bm{g}({\bm{x},\bm{y}})\label{algebraic eqn}
\end{eqnarray}
Equation (\ref{fast ode}) describes generator dynamics, and their associated control; (\ref{algebraic eqn}) describes the electrical transmission system and the static behaviors of devices. $\bm{f}$ and $\bm{g}$ are continuous functions, vectors $\bm{x}\in\mathbb{R}^{n_{\bm{x}}}$ and $\bm{y}\in\mathbb{R}^{n_{\bm{y}}}$ are the corresponding state variables (generator rotor angles, rotor speeds) and algebraic variables (bus voltages, bus angles) \cite{Wangxz:CAS}.

In this paper, we focus on ambient oscillations around stable steady state. Hence, we demonstrate the proposed method using the classical generator model, which can typically represent generator dynamics in ambient conditions. We assume that the load variations and renewable injections can be transformed into the variation of generator mechanical power, 
i.e., the mechanical power for Generator $i$ is $P_{m_i}+\sigma_i\xi_i$, where $\xi_i$ is a standard Gaussian noise, and $\sigma_i^2$ is the noise variance.
Thus (\ref{fast ode})-(\ref{algebraic eqn}) can be represented as:
\begin{eqnarray}
\dot{\bm{\delta}}&=&\bm{\omega}\label{swing-1}\\
M\dot{\bm{\omega}}&=&\bm{P_m}-\bm{P_e}-{D}\bm{\omega}+{\Sigma}\bm{\xi}\label{swing-2}
\end{eqnarray}
with
\begin{equation}
P_{e_i}=\sum_{j=1}^{n}E_iE_j(G_{ij}\cos({\delta}_i-{\delta}_j)+B_{ij}\sin(\delta_i-{\delta}_j))\label{swing-3}
\end{equation}
Particularly, $\bm{\delta}=[\delta_1,...\delta_n]^T$ is a vector of generator rotor angles, $\bm{\omega}=[\omega_1,...\omega_n]^T$ is a vector of generator rotor speeds, $\bm{P_m}=[P_{m_1},...P_{m_n}]^T$ is a vector of generators' mechanical power, $\bm{P_e}=[P_{e_1},...P_{e_2}]^T$ is a vector of generators' electrical power, $M=\mbox{diag}(M_1,...M_n)$ are the inertia constants, $D=\mbox{diag}(D_1,...D_n)$ are the damping factors. In addition, $\bm{{\xi}}$ is a vector of independent standard Gaussian random variables representing the variation of power injections, and $\Sigma=\mbox{diag}(\sigma_1,...\sigma_n)$ is the covariance matrix.
For the sake of simplicity, in this work we model the loads as constant impedances. In the future, efforts are needed to incorporate more realistic models.


Linearizing (\ref{swing-1})-(\ref{swing-2}) gives the following:
\begin{eqnarray}
\left[\begin{array}{c}\dot{\bm{\delta}}\\\dot{\bm{\omega}}\end{array}\right]&=&
\left[\begin{array}{cc}{{0}}&{I_n}\\-M^{-1}\frac{\partial{\bm{P_e}}}{\partial{\bm{\delta}}}&-M^{-1}D\end{array}\right]
\left[\begin{array}{c}{\bm{\delta}}\\{\bm{\omega}}\end{array}\right]\nonumber\\
&&+\left[\begin{array}{c}0\\M^{-1}\Sigma\end{array}\right]\bm{\xi}\label{swing-matrix}
\end{eqnarray}
Let $\bm{x}=[\bm{\delta},\bm{\omega}]^T$, $A=\left[\begin{array}{cc}{{0}}&{I_n}\\-M^{-1}\frac{\partial{\bm{P_e}}}{\partial{\bm{\delta}}}&-M^{-1}D\end{array}\right]$, $B=[0,M^{-1}\Sigma]^T$, then (\ref{swing-matrix}) takes the form:
\begin{equation}
\dot{\bm{x}}=A\bm{x}+B\bm{\xi}
\end{equation}
Specifically, if state matrix $A$ is stable, the stationary covariance matrix $C_{\bm{x}\bm{x}}=\left[\begin{array}{cc}C_{\bm{\delta}{\bm{\delta}}}&C_{\bm{\delta}{\bm{\omega}}}\\C_{\bm{\omega}{\bm{\delta}}}&C_{\bm{\omega}{\bm{\omega}}}\end{array}\right]$ can be shown to satisfy the following Lyapunov equation \cite{Hines:2015}\cite{Gardiner:2009}:
\begin{equation}
AC_{\bm{x}\bm{x}}+C_{\bm{x}\bm{x}}A^T=-BB^T \label{lyapunov}
\end{equation}
which nicely combine the measurement and the model knowledge.
In this paper, we draw upon this relation to estimate the state matrix $A$ from the statistical properties of state $C_{\bm{x}\bm{x}}$ that can be extracted from PMU measurements.

Substituting the detailed expressions of $A$ and $B$ to (\ref{lyapunov}) and performing algebraic simplification, we have that:
\begin{eqnarray}
C_{\bm{\delta}{\bm{\omega}}}&=&0\\
C_{\bm{\delta}{\bm{\delta}}}&=&(\frac{\partial{\bm{P_e}}}{\partial{\bm{\delta}}})^{-1}MC_{\bm{\omega}{\bm{\omega}}}\label{rdd}\\
C_{\bm{\omega}{\bm{\omega}}}&=&\frac{1}{2}M^{-1}D^{-1}\Sigma^2\label{rww}
\end{eqnarray}
Particularly, we utilize the relation (\ref{rdd}) that combines the measurements of states and the physical model, which provides an ingenious way to estimate the dynamic state Jacobian matrix from the measurements.
Given that the inertia constants $M$ are typically known, and $C_{\bm{\delta}\bm{\delta}}$, $C_{\bm{\omega}{\bm{\omega}}}$ can be extracted from the PMU measurements (see details in Section \ref{subsectioncddcww}), the Jacobian matrix $\frac{\partial{\bm{P_e}}}{\partial{\bm{\delta}}}$ can be computed from (\ref{rdd}). Furthermore, the system state matrix $A$ can be readily constructed if the damping constants $D$ are known. 

\subsection{Estimating Covariance Matrixes $C_{\bm{\delta}\bm{\delta}}$ and $C_{\bm{\omega}{\bm{\omega}}}$}
We assume that PMUs are installed at all the substations that generators are connected to, and that we can use the PMUs to calculate the rotor angle $\bm{\delta}$ and rotor speed $\bm{\omega}$ in steady state with ambient oscillations. Discussion how exactly it is done is beyond the scope of this paper and we refer the
reader to \cite{Zhou:2011}-\cite{Liu:2011}.
\label{subsectioncddcww}

By definition
\begin{equation}\label{cdd}
  C_{\bm{\delta}\bm{\delta}}=\left[\begin{array}{cccc} C_{\delta_1\delta_1}&C_{\delta_1\delta_2}&\dots&C_{\delta_1\delta_n}\\
  C_{\delta_2\delta_1}&C_{\delta_2\delta_2}&\dots&C_{\delta_2\delta_n}\\
  \vdots&\vdots&\ddots&\vdots\\
  C_{\delta_n\delta_1}&C_{\delta_n\delta_2}&\dots&C_{\delta_n\delta_n}
  \end{array}\right]
\end{equation}
where $C_{\delta_i\delta_j}=\E[(\delta_i-\mu_i)(\delta_j-\mu_j)]$, and $\mu_i$ is the mean of $\delta_i$.
However, $C_{\bm{\delta}\bm{\delta}}$ is typically unknown in practice and needs to be estimated from limited PMU data. A window size around $300$s is implemented in the examples of this paper, which shows a good accuracy. An unbiased estimator of $C_{\bm{\delta}\bm{\delta}}$ is the sample covariance matrix $Q_{\bm{\delta}\bm{\delta}}$ each entry of which is calculated as below\cite{Gardiner:2009}:
\begin{equation}
Q_{\delta_i \delta_j}=\frac{1}{N-1}\sum_{k=1}^N(\delta_{ki}-\bar{\delta}_i)((\delta_{kj}-\bar{\delta}_j))\label{qdd}
\end{equation}
where $\bar{\delta}_i$ is the sample mean of $\delta_i$, and $N$ is the sample size. Likewise, $C_{\bm{\omega}\bm{\omega}}$ can be estimated by $Q_{\bm{\omega}\bm{\omega}}$ in the same way:
\begin{equation}
Q_{\omega_i \omega_j}=\frac{1}{N-1}\sum_{k=1}^N(\omega_{ki}-\bar{\omega}_i)((\omega_{kj}-\bar{\omega}_j))\label{qww}
\end{equation}

When $Q_{\bm{\delta}\bm{\delta}}$ and $Q_{\bm{\omega}\bm{\omega}}$ are calculated, and with the parameter values $M$ on file, we are able to calculate the Jacobian matrix $\frac{\partial{\bm{P_e}}}{\partial{\bm{\delta}}}$ from (\ref{rdd}):
\begin{equation}
(\frac{\partial\bm{P_e}}{\partial\bm{{\delta}}})=MQ_{\bm{{\omega}}{\bm{{\omega}}}}Q^{-1}_{\bm{{\delta}}{\bm{{\delta}}}}\label{approxjacobian}
\end{equation}



\section{Numerical Illustration}
In this section, a numerical example is presented to show the validity of the proposed method. In addition, it will be shown that the proposed method may help identify big discrepancy in the assumed network model since its performance is robust to network topology change.

We consider the standard WSCC 3-generator, 9-bus system model (see, e.g. \cite{Chiang:book}). 
The system model in the center-of-inertia (COI) formulation is given as below:
\begin{eqnarray}
\dot{\tilde{\delta}}_1&=&\tilde{\omega}_1\label{9bus-1}\\
\dot{\tilde{\delta}}_2&=&\tilde{\omega}_2\\
M_1\dot{\tilde{\omega}}_1&=&P_{m_1}-P_{e_1}-\frac{M_1}{M_T}P_{coi}-D_1\tilde{\omega}_1+\sigma_1\xi_1\\
M_2\dot{\tilde{\omega}}_2&=&P_{m_2}-P_{e_2}-\frac{M_2}{M_T}P_{coi}-D_2\tilde{\omega}_2+\sigma_2\xi_2\label{9bus-2}
\end{eqnarray}
where $\delta_0=\frac{1}{M_T}\sum_{i=1}^{3}M_i\delta_i$, $\omega_0=\frac{1}{M_T}\sum_{i=1}^{3}M_i\omega_i$, $M_T=\sum_{i=1}^{3}M_i$, $\tilde{\delta}_i=\delta_i-\delta_0$, $\tilde{\omega}_i=\omega_i-\omega_0$, for $i=1,2,3$,
and
\begin{eqnarray}
P_{e_i}&=&\sum_{j=1}^{3}E_iE_j(G_{ij}\cos(\tilde{\delta}_i-\tilde{\delta}_j)+B_{ij}\sin(\tilde{\delta}_i-\tilde{\delta}_j))\nonumber\\
P_{coi}&=&\sum_{i=1}^{3}(P_{m_i}-P_{e_i})
\end{eqnarray}
The parameter values in this examples are: $P_{m_1}=0.72$ p.u., $P_{m_2}=1.63$ p.u., $P_{m_3}=0.85$ p.u.; $E_1=1.057$ p.u., $E_2=1.050$ p.u., $E_3=1.017$ p.u; $M_1=0.63$, $M_2=0.34$, $M_3=0.16$; $D_1=0.63$, $D_2=0.34$, $D_3=0.16$.
Because the following relations that $\tilde{\delta}_3=-\frac{M_1\tilde{\delta}_1+M_2\tilde{\delta}_2}{M_3}$ and $\tilde{\omega}_3=-\frac{M_1\tilde{\omega}_1+M_2\tilde{\omega}_2}{M_3}$ hold in the COI formulation, $\tilde{\delta}_3$ and $\tilde{\omega}_3$ depending on the other state variables can be obtained without integration.

The system state matrix is as follows:
\begin{equation}
A=\left[\begin{array}{cc|cc}0&0&1&0\\0&0&0&1\\\hline\multicolumn{2}{c}{\multirow{2}{*}{J}}\vline\hspace{-0.002in}&-\frac{D_1}{M_1}&0\\&&0&-\frac{D_2}{M_2}\end{array}\right]\label{A}
\end{equation}
where $J=-M^{-1}(\frac{\partial\bm{P_e}}{\partial\bm{\tilde{\delta}}}+M\frac{1}{M_T}\frac{\partial P_{coi}}{\partial\bm{\tilde{\delta}}})$, for $i=1,2$. Let $(\frac{\partial\bm{P_e}}{\partial\bm{\tilde{\delta}}})_{coi}=\frac{\partial\bm{P_e}}{\partial\bm{\tilde{\delta}}}+M\frac{1}{M_T}\frac{\partial P_{coi}}{\partial\bm{\tilde{\delta}}}$, then we have
\begin{equation}
\small{((\frac{\partial\bm{P_e}}{\partial\bm{\tilde{\delta}}})_{coi})_{ij}}=\left\{\begin{array}{l}E_iE_j(G_{ij}\sin(\tilde{\delta}_i-\tilde{\delta}_j)-B_{ij}\cos(\tilde{\delta}_i-\tilde{\delta}_j))\\
+\frac{M_i}{M_T}\frac{\partial P_{coi}}{\partial\tilde{\delta_i}} \hspace{1.2in} \mbox{if $i\not=j$}\\
\sum^{n}_{k=1}E_iE_k(G_{ik}\sin(\tilde{\delta}_i-\tilde{\delta}_k)\\
+B_{ik}\cos(\tilde{\delta}_i-\tilde{\delta}_k))+\frac{M_i}{M_T}\frac{\partial P_{coi}}{\partial\tilde{\delta_i}} \hspace{0.1in} \mbox{if $i=j$}
\end{array}\right.\label{dpeddcoi}
\end{equation}
where $\frac{\partial P_{coi}}{\partial \tilde{\delta_i}}=2\sum_{k\not=i}E_iE_kG_{ik}\sin(\tilde{\delta}_i-\tilde{\delta}_k)$.

If the network model as well as system states are available, the Jacobian matrix $(\frac{\partial\bm{P_e}}{\partial\bm{\tilde{\delta}}})_{coi}$ can be directly computed from (\ref{dpeddcoi}). However, the system topology and line model parameter values are subjected to continuous perturbations. Therefore, the exact knowledge of network topology with up-to-date network parameter values may not be available. In addition, the control faults and transmission delays may also lead to imprecise knowledge of the network parameter values.

\begin{figure}[!ht]
\centering
\begin{subfigure}[t]{0.52\linewidth}
\includegraphics[width=1.8in ,keepaspectratio=true,angle=0]{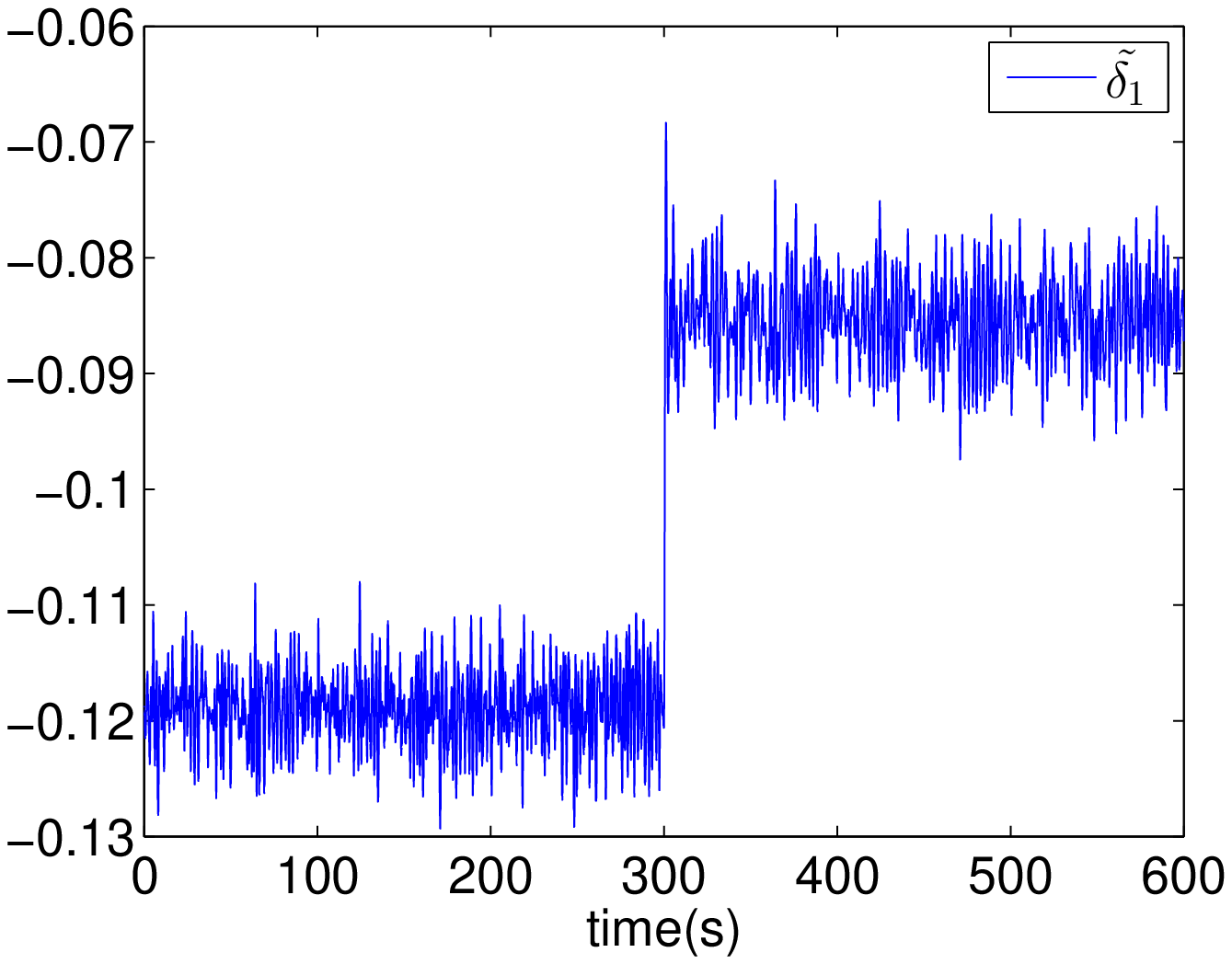}
\caption{Trajectory of $\tilde{\delta}_1$ on [0s,600s]}\label{d1-9}
\end{subfigure}%
\begin{subfigure}[t]{0.5\linewidth}
\includegraphics[width=1.8in ,keepaspectratio=true,angle=0]{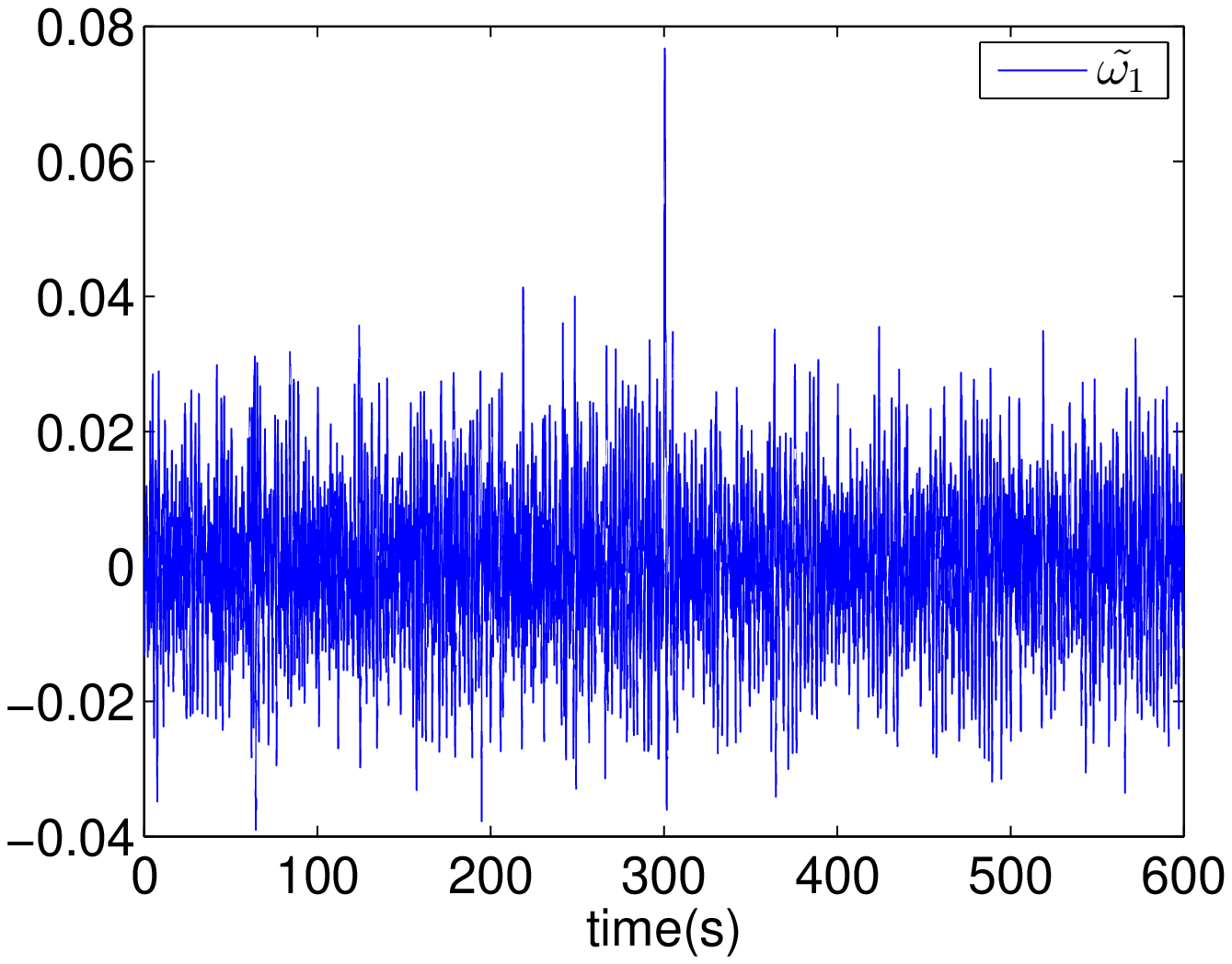}
\caption{Trajectory of $\tilde{\omega}_1$ on [0s,600s]}\label{w1-9}
\end{subfigure}
\caption{Trajectories of the state variables in the 9-bus system in COI reference.}\label{9bus}
\end{figure}

In contrast, the proposed method does not require the knowledge of network parameters. 
In order to show this, we conduct the following numerical experiment. Assuming that the transient reactance $x_d'$ of Generator 1 increases from $0.0608$ p.u. to $0.1824$ p.u. at 300.01s, mimicking a ling loss between the generator internal node and its terminal bus \cite{Pai:2012}. Let $\sigma_1=\sigma_2=0.01$, 
the trajectories of some state variables in system (\ref{9bus-1})-(\ref{9bus-2})
before and after the contingency are presented in Fig. \ref{9bus}, 
from which we see that the system is able to maintain stability after the line loss, and the state variables are always fluctuating around the stable steady states due to the variation of load and generator power.

Before the contingency and if there is no stochastic variation, i.e., $\sigma_1=\sigma_2=0$, $(\frac{\partial\bm{P_e}}{\partial\bm{\tilde{\delta}}})_{coi}$ is a constant matrix that can be easily acquired from (\ref{dpeddcoi}):
\begin{equation}
(\frac{\partial\bm{P_e}}{\partial\bm{\tilde{\delta}}})_{coi}=\left[ \begin{array}{cc} 8.053 & 1.240\\ 2.802 & 5.085\end{array}\right]\label{det_before}
\end{equation}
We want to show that the matrix obtained from the proposed method is close to this model-based deterministic matrix.

First, $Q_{\bm{\tilde{\omega}}{\bm{\tilde{\omega}}}}$ and $Q_{\bm{\tilde{\delta}}{\bm{\tilde{\delta}}}}$ before the contingency can be calculated from the system trajectories on [0s, 300s]:
\begin{eqnarray}
Q_{\bm{\tilde{\delta}}{\bm{\tilde{\delta}}}}&=&10^{-4}\times\left[ \begin{array}{cc}  0.106 & -0.0483\\ -0.0483 &0.359\end{array}\right]\nonumber\\
C_{\bm{\tilde{\omega}}{\bm{\tilde{\omega}}}}&=&10^{-3}\times\left[ \begin{array}{cc}  0.123 & 0.008\\ 0.008 &0.514\end{array}\right]\nonumber
\end{eqnarray}
and therefore $(\frac{\partial\bm{P_e}}{\partial\bm{\tilde{\delta}}})_{coi}$ can be computed by the proposed method according to (\ref{approxjacobian}):
\begin{equation}
(\frac{\partial\bm{P_e}}{\partial\bm{\tilde{\delta}}})_{coi}^{\star}=\left[ \begin{array}{cc} 7.806 & 1.192\\  2.642 &  5.214 \end{array}\right]
\end{equation}
where $^\star$ denotes the Jacobian matrix estimated by the proposed method. It is seen that $(\frac{\partial\bm{P_e}}{\partial\bm{\tilde{\delta}}})_{coi}^{\star}$ and $(\frac{\partial\bm{P_e}}{\partial\bm{\tilde{\delta}}})_{coi}$ are close to each other. Specifically, the estimation error is:
\begin{equation}
\frac{\|(\frac{\partial\bm{P_e}}{\partial\bm{\tilde{\delta}}})_{coi}^\star-(\frac{\partial\bm{P_e}}{\partial\bm{\tilde{\delta}}})_{coi}\|_F}{\|(\frac{\partial\bm{P_e}}{\partial\bm{\tilde{\delta}}})_{coi}\|_F}=3.25\%\label{matrixdistance} \end{equation}
where $\|\|_F$ denotes the Frobenius norm of a matrix measuring the distance between two matrixes. Assuming the damping constants $D$ are known, we can also compute the system state matrix $A$ and the resulting estimation error is:
\begin{equation}
\frac{\|A^\star-A\|_F}{\|A\|_F}=0.34\%\label{matrixdistance}
\end{equation}
The above results demonstrate the proposed method is able to provide accurate estimation for the dynamic state Jacobian matrix and the system state matrix.

To highlight the value of the proposed hybrid method, we assume that the topology change is undetected, while the change of  nominal states of $\delta$ and $\omega$ can be detected via PMU measurements. Therefore, the Jacobian matrix after the contingency obtained from the model-based method by (\ref{dpeddcoi}) is:
\begin{equation}
\overline{(\frac{\partial\bm{P_e}}{\partial\bm{\tilde{\delta}}})_{coi}^\diamond}=\left[ \begin{array}{cc} 8.171 & 1.239\\  2.810 &  5.150 \end{array}\right]\label{modelbased}
\end{equation}
where the overline denotes the values after the contingency, and $^\diamond$ denotes the value obtained from the model-based method. Indeed, this estimated Jacobian matrix is far away from the true value of the
Jacobian matrix after the contingency shown as below:
\begin{equation}
\overline{(\frac{\partial\bm{P_e}}{\partial\bm{\tilde{\delta}}})_{coi}}=\left[ \begin{array}{cc} 5.950 & 0.968\\  3.885 &  5.168\end{array}\right]\label{true_after}
\end{equation}
due to the out-of-date network parameter values.

In contrast, by applying the proposed method, we obtain that:
\begin{eqnarray}
\overline{Q_{\bm{\tilde{\delta}}{\bm{\tilde{\delta}}}}}&=&10^{-4}\times\left[ \begin{array}{cc}  0.139 & -0.108\\ -0.108 &0.495\end{array}\right]\nonumber\\
\overline{Q_{\bm{\tilde{\omega}}{\bm{\tilde{\omega}}}}}&=&10^{-3}\times\left[ \begin{array}{cc}  0.113 & -0.024\\ -0.024 &0.658\end{array}\right]\nonumber\\
\overline{(\frac{\partial\bm{P_e}}{\partial\bm{\tilde{\delta}}})_{coi}^{\star}}&=&\left[ \begin{array}{cc} 5.853 & 0.976\\ 3.524 & 5.289 \end{array}\right]\label{proposed}
\end{eqnarray}


The Frobenius distance between the true (\ref{true_after}) and the estimated Jacobian matrix by the proposed method (\ref{proposed}) is still small and is equal to 4.45\%. However, the distance between the true (\ref{true_after}) and the model-based estimation (\ref{modelbased}) is equal to 28.13\% due to assumed inaccurate network model. Regarding the system state matrix $A$, the similar distances are 5.33\% and 22.37\%.
These results clearly demonstrate that the proposed hybrid method provides more accurate estimation for the Jacobian matrix after the contingency since its performance is not affected by the change of network topology. From the other hand, The big difference between the model-based (\ref{modelbased}) and the measurement-based matrix (\ref{proposed}) indicates that there was a mistake in the assumed system model that needs great attention.


From this example, some important insights can be obtained. The proposed method is able to provide accurate estimation for the dynamic state Jacobian matrix 
by exploiting the statistical properties of the stochastic system. In addition, the performance of the proposed method may outstand under imprecise knowledge or undetectable change of network topology. The big difference between model-based and the proposed estimations may also alarm system operators for an assumed inaccurate system model.

\section{conclusions and perspectives}\label{sectionconclusion}

In this paper, we have proposed a hybrid measurement and model-based method for estimating dynamic state Jacobian matrix in near real-time. The proposed hybrid method works as a grey box bridging the measurement and the model, and is able to provide fairly accurate estimation without being affected by the variation of network topology. In addition, the proposed method may also identify big discrepancy in the assumed network model.


Since the dynamic state Jacobian matrix and the system state matrix carry uttermost important information on system conditions and dynamics, they can be utilized in various applications such as online oscillation analysis, stability monitoring and emergency control, congestion relief and so forth. In the future, we plan to explore these applications of the estimated Jacobian matrix in power system operation. Besides, further investigations of the method on higher-order generator models and detailed load models are expected. 

\end{document}